\documentclass[twocolumn,pra,aps]{revtex4} 
\usepackage{epsfig,amsmath}
\usepackage{graphicx}
\usepackage{dcolumn}
\usepackage{bm}
\usepackage{bbold}
\newcommand{\beq}{\begin{equation}}
\newcommand{\eeq}{\end{equation}}
\newcommand{\beqa}{\begin{eqnarray}}
\newcommand{\eeqa}{\end{eqnarray}}

\begin{document}
\author{Yi Ding}
\email{yding2020swjtu@163.com}
\affiliation{School of Physical Science and Technology, Southwest Jiaotong University, Chengdu 610031, China}
\title{Scattering of light waves from a collection with particles of $\mathcal{L}$ types}
\date{\today}

\begin{abstract}
A new approach is developed within the first-order Born approximation to light scattering from a collection of particles with $\mathcal{L}$ types. Two $\mathcal{L}\times\mathcal{L}$ matrices called pair-potential matrix (PPM) and pair-structure matrix (PSM) are introduced to jointly formulate the coherence properties of the scattered field. We derive a closed-form relation that associates the cross-spectral density function of the scattered field with the PPM and the PSM, showing that the the cross-spectral density function equals the trace of the product of the PSM and the transpose of the PPM. Based on this, the spectral degree of coherence (SDOC) of the scattered field is further analysed. We show that for a special case where the spatial distributions of scattering potentials of different types of particles are similar and the same is true of their density distributions, the PPM and the PSM will reduce to two new matrices whose elements separately quantify degree of angular correlation of the scattering potentials of particles and their density distributions, and the number of species of particles in this special case as a scaled factor ensures the normalization of the SDOC. Two special hybrid particulate systems as examples are given to illustrate the importance of our new approach.

\end{abstract}

\maketitle
\section{Introduction}
As one of the most common optical phenomena in nature, light scattering from collections of particles is always of interest in many fields such as optical trapping, optical imaging, biomedical diagnostics and atmospheric optics, to name a few. There have been many efforts to explore light scattering from particulate media \cite{DW, Gbur, Skoro, Skoro1, TONG, Tong, Olga, MEI, DZ, WWZ, Peng, Dingz, OB, Wang}, especially from a collection in which the scattering potentials of individual particles are deterministic functions of position, but the locations of particles vary randomly in space \cite{DW, Gbur, Tong, WWZ, Peng, Dingz, OB}. The collective properties of the system in this situation are determined by the so-called pair-structure factor \cite{Skoro}. It has been shown that, in addition to the scattering potentials of individual particles, the knowledge of the pair-structure factor is sufficient for determining all the second-order statistical properties of fields produced on scattering from collections of particles. However, the further study has shown that even if the pair-structure factor is known, it is still not adequate to describe light scattering from a collection of particles with $\mathcal{L}$ types \cite{Tong}. Another important quantity called joint pair-structure factor needs to be introduced as a measure of the correlation property of between particles across different types. There actually exists a $\mathcal{L}\times\mathcal{L}$ matrix whose diagonal elements are the pair-structure factors of each particle type and off-diagonal elements are the joint pair-structure factors for each pair of particle types. Since the entire matrix contains all the information between the correlation properties of between particles within one type and across different types, it is called pair-structure matrix (PSM). Regarding the scattering potentials of particles, they were arranged to form a column vector, and finally the cross-spectral density function of the scattered field was written as a product of the Hermitian adjoint of the column vector and the pair-structure matrix as well as the column vector itself. 

Although such a representation on the cross-spectral density function of the scattered field greatly simplifies the mathematical complexity of the scattered field, it is not universal. Once the randomness of the scattering potentials of individual particles in the collection is initiated, they cannot be arranged to form a column vector, leading the representation of the cross-spectral density function to invalidate. Actually, there is another $\mathcal{L}\times\mathcal{L}$ matrix hidden, which is called pair-potential matrix (PPM) characterizing angular correlations of scattering potentials of particles of the same and of different types. It can be utilized, together with the PSM, to characterize the scattered field generated by collections of particles with $\mathcal{L}$ types.  Both these two matrices play an indispensable role in the description of the weak scattering process of light waves from a hybrid particulate system, and neither of them is better than the other.

In this work, we will develop a new approach to light scattering from a collection of particles with $\mathcal{L}$ types, based on the PPM and the PSM. We will formulate a closed-form relation that associates the cross-spectral density function of the scattered field with the PPM and the PSM, and thus it is sufficient to determine all the second-order statistical properties of the scattered fields from these two matrices. Our new approach is quite general and can largely simplify the theoretical procedures of the scattering of light from complex collection of scatters. 

\section{Convolution representation of the scattering potential and its correlation function of a collection of particles with $\mathcal{L}$ types}
For a collection of particles, there are usually $\mathcal{L}$ types of particles forming this system, $m(p)$ of each type, ($p=1,2,3,\cdot\cdot\cdot,L$), located at points specified by position vectors $\mathbf{r}_{pm}$. We characterize the response of each particle to an incoming field by a scattering potential $f_{p}(\mathbf{r}^{\prime})$, which is closely related to the refractive index of the particle. The scattering potential $F(\mathbf{r}^{\prime},\omega)$ of the whole collection can be usually defined as \cite{DW} 
\begin{align}
    F(\mathbf{r}^{\prime},\omega)=\sum_{p}^{\mathcal{L}}\sum_{m(p)}f_{p}(\mathbf{r}^{\prime}-\mathbf{r}_{pm},\omega).
\end{align}
For the sake of following discussions, we now rewrite the definition of the scattering potential of the collection in a slightly unfamiliar form, viz., 
\begin{equation}\label{scatteringpotential}
    F(\mathbf{r}^{\prime},\omega)=\sum_{p=1}^{\mathcal{L}}f_{p}(\mathbf{r}^{\prime},\omega)\otimes g_{p}(\mathbf{r}^{\prime}),
\end{equation}
where 
\begin{equation}\label{densityfunction}
    g_{p}(\mathbf{r}^{\prime})\equiv\sum_{m(p)}\delta(\mathbf{r}^{\prime}-\mathbf{r}_{pm})
\end{equation}
may be interpreted as the density function of the $p$th-type particle \cite{Gbur}. This is natural if the $p$th-type particle in the collection can be effectively regarded as consisting of $m(p)$ `point particle'. $\delta(\cdots)$ is the three-dimensional Dirac delta function, and $\otimes$ denotes the convolution operation.

In general, the collection may be of deterministic or random nature. In the case when the collection is deterministic, its scattering potential $F(\mathbf{r}^{\prime},\omega)$ is a well-defined function of position. However, for a more involved case when the scattering potential of the collection is not deterministic, but varies randomly as a function of position. In this case, the spatial correlation function of scattering potential, specified by position vectors $\mathbf{r_{1}}^{\prime}$ and $\mathbf{r_{2}}^{\prime}$, may be given as (\cite{Wolf2}, Sec. 6.3.1)

\begin{equation}\label{correlationfunction}
{C_{F}}\left( {{{\mathbf{{r_{1}^{\prime}}}}},{{\mathbf{{r_{2}^{\prime}}}}}%
,\omega}\right) =\left\langle {{F^{\ast}}\left( {{{\mathbf{{r_{1}^{\prime}%
}}}},\omega}\right) F\left( {{{\mathbf{{r_{2}^{\prime}}}}},\omega}\right)
}\right\rangle_{m},
\end{equation}
where $\left\langle\cdots\right\rangle_{m}$ stands for the average taken over different realizations of the scatterer. On substituting from Eq. \eqref{scatteringpotential} into Eq. \eqref{correlationfunction}, after some simple rearrangements, we end up with

\begin{equation}\label{sc}
 {C_{F}}\left( {{{\mathbf{{r_{1}^{\prime}}}}},{{\mathbf{{r_{2}^{\prime}}}}}%
,\omega}\right)=\sum_{p=1}^{\mathcal{L}}\sum_{q=1}^{\mathcal{L}}{C_{f_{pq}}}\left({{{\mathbf{{r_{1}^{\prime}}}}},{{\mathbf{{r_{2}^{\prime}}}}}%
,\omega}\right)\otimes{C_{g_{pq}}}\left({{{\mathbf{{r_{1}^{\prime}}}}},{{\mathbf{{r_{2}^{\prime}}}}}%
}\right).
\end{equation}
where 

\begin{equation}\label{scatteringcorrelation}
   {C_{f_{pq}}}\left({{{\mathbf{{r_{1}^{\prime}}}}},{{\mathbf{{r_{2}^{\prime}}}}},\omega}\right)=\left\langle f_{p}^{*}(\mathbf{r}_{1}^{\prime},\omega)f_{q}(\mathbf{r}_{2}^{\prime},\omega)\right\rangle 
\end{equation}
represent the self-correlation functions of the scattering potentials of particles of same type (if $p=q$) or the cross-correlation functions of the scattering potentials of particles of different types (if $p\neq q$), and 

\begin{equation} \label{densitycorrelation}
{C_{g_{pq}}}\left({{{\mathbf{{r_{1}^{\prime}}}}},{{\mathbf{{r_{2}^{\prime}}}}}}\right)=\left\langle g_{p}^{*}(\mathbf{r}_{1}^{\prime})g_{q}(\mathbf{r}_{2}^{\prime})\right\rangle    
\end{equation}
represent the self-correlation functions of the density functions of particles of same type (if $p=q$) or the cross-correlation functions of the density functions of particles of different types (if $p\neq q$).

It is seen in the transition from Eqs. \eqref{scatteringpotential} and \eqref{correlationfunction} to Eq. \eqref{sc} that use has been made of the assumption that the average over the ensemble of the scattering potentials of particles and that over the ensemble of their density distributions are mutually independent. 

\section{relation between the cross-spectral density function of the scattered field and the PPM and the PSM}

\begin{figure}[bthp]
\centering
\includegraphics[width=8cm]{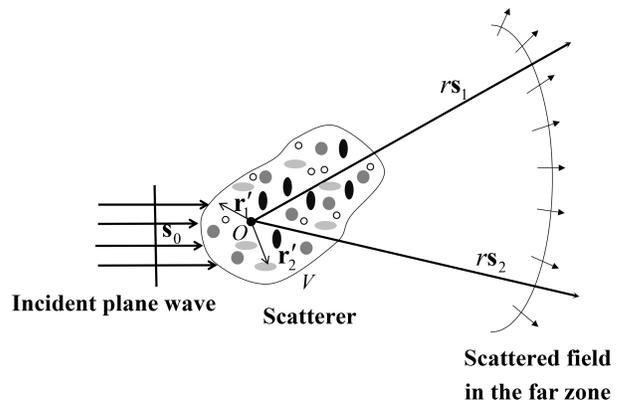}
\caption{Illustration of notations.}
\label{Fig 1}
\end{figure}

Assume now that a coherent polychromatic plane light wave, propagating in a direction specified by a real unit vector $\mathbf{s}_{0}$, is incident upon a statistically stationary a particulate medium (see Fig. \ref{Fig 1}), occupying a finite domain $\mathcal{V}$. The statistical property of the incident field at a pair of points $\mathbf{r_{1}}^{\prime}$ and $\mathbf{r_{2}}^{\prime}$ within the domain of the collection can be characterized by its cross-spectral density function with a form of
\begin{equation}\label{element}
  W^{\text{(i)} }(\mathbf{r_{1}}^{\prime},\mathbf{r_{2}}^{\prime},\mathbf{s}_{0};\omega)=S^{\text{(in)}}(\omega)\exp[ik\mathbf{s_{0}} \cdot (\mathbf{r}_{1}^{\prime}-\mathbf{r}_{2}^{\prime})],
\end{equation}
where $S^{\text{(in)}}(\omega)$ is the spectrum of the incident field, and $k =\omega/c$ is the wave number with $c$ being the speed of light in vacuum and $\omega$ being the angular frequency. 

It is well known that within the validity of the first-order Born approximation, the cross-spectral density function of the scattered field at two points specified by position vectors $r\mathbf{s_{1}}$ and $r\mathbf{s_{2}}$ in the far-zone out of scatterer can be formulated as (\cite{Wolf2}, Sec. 6.3)
\begin{align}\label{crossspectraldensity}
    W^{(\text{s})}(r\mathbf{s}_{1},r\mathbf{s}_{2},\omega)=\frac{S^{\text{(in)}}(\omega)}{r^{2}}\widetilde{C}_{F}\Bigl[-\mathbf{K}_{1},\mathbf{K}_{1},\omega\Bigr],
\end{align}
where
\begin{align}\label{1}
  \widetilde{C}_{F}(\mathbf{K}_{1},\mathbf{K}_{2},\omega)&=\int_{\mathcal{V}}\int_{\mathcal{V}}{C_{F}}\left({{{\mathbf{{r_{1}^{\prime}}}}},{{\mathbf{{r_{2}^{\prime}}}}}%
,\omega}\right) \notag \\ & \times \exp\Bigl[{-i(\mathbf{K}_{2}\cdot{r}_{2}^{\prime}+\mathbf{K}_{1}\cdot{r}_{1}^{\prime})}\Bigr]d^{3}r_{1}^{\prime}d^{3}r_{2}^{\prime}
\end{align}
is the six-dimensional spatial Fourier transformation of the correlation function of the scattering potential of the scatter, and $\mathbf{K_{1}}=k(\mathbf{s}_{1}-\mathbf{s}_{0})$ and $\mathbf{K_{2}}=k(\mathbf{s}_{2}-\mathbf{s}_{0})$ are analogous to the momentum transfer vector of quantum mechanical theory of potential scattering (\cite{Wolf2}, Sec. 6.1).

On substituting from Eq. \eqref{sc} into Eq. \eqref{1}, using the mathematical theorem that the Fourier transform of the convolution of two functions is a product of their Fourier transforms, we get 
\begin{align}\label{crossspectraldensity1}
   \widetilde{C}_{F}(\mathbf{K}_{1},\mathbf{K}_{2},\omega)&=\sum_{p=1}^{\mathcal{L}}\sum_{q=1}^{\mathcal{L}}\widetilde{C}_{f_{pq}}(-\mathbf{K}_{1},\mathbf{K}_{2},\omega)\notag \\
    &\times\widetilde{C}_{g_{pq}}(-\mathbf{K}_{1},\mathbf{K}_{2},\omega),
\end{align}
where 
\begin{align}\label{2}
  \widetilde{C}_{f_{pq}}(\mathbf{K}_{1},\mathbf{K}_{2},\omega)&=\int_{\mathcal{V}}\int_{\mathcal{V}}{C_{f_{pq}}}\left({{{\mathbf{{r_{1}^{\prime}}}}},{{\mathbf{{r_{2}^{\prime}}}}}%
,\omega}\right) \notag \\ & \times \exp\Bigl[{-i(\mathbf{K}_{2}\cdot{r}_{2}^{\prime}+\mathbf{K}_{1}\cdot{r}_{1}^{\prime})}\Bigr]d^{3}r_{1}^{\prime}d^{3}r_{2}^{\prime}
\end{align}
are the six-dimensional spatial Fourier transformations of the self-correlation functions of the scattering potentials of particles of same type (if $p=q$) or the cross-correlation functions of the scattering potentials of particles of different types (if $p\neq q$), and 
\begin{align}\label{3}
  \widetilde{C}_{g_{pq}}(\mathbf{K}_{1},\mathbf{K}_{2},\omega)&=\int_{\mathcal{V}}\int_{\mathcal{V}}{C_{g_{pq}}}\left({{{\mathbf{{r_{1}^{\prime}}}}},{{\mathbf{{r_{2}^{\prime}}}}}%
,\omega}\right) \notag \\ & \times \exp\Bigl[{-i(\mathbf{K}_{2}\cdot{r}_{2}^{\prime}+\mathbf{K}_{1}\cdot{r}_{1}^{\prime})}\Bigr]d^{3}r_{1}^{\prime}d^{3}r_{2}^{\prime}
\end{align}
are the six-dimensional spatial Fourier transformations of the self-correlation functions of the density distributions of particles of same type (if $p=q$) or the cross-correlation functions of the density distributions of particles of different types (if $p\neq q$).

We now introduce two $\mathcal{L}\times \mathcal{L}$ matrices to jointly formulate the scattered field. The first matrix is defined as 
\begin{align}\label{matrix1}
    \mathcal{F}(\mathbf{K_{1}},\mathbf{K_{2}},\omega)&=\Bigl[\widetilde{C}_{f_{pq}}(-\mathbf{K_{1}},\mathbf{K_{2}},\omega)\Bigr]_{\mathcal{L}\times\mathcal{L}}.
\end{align}
From Eqs. \eqref{scatteringcorrelation} and \eqref{2}, it readily follows that the diagonal elements of this matrix represent angular self-correlations of the scattering potentials of particles of same type and the off-diagonal elements represent angular cross-correlations of the scattering potentials of each pair of particle types. The entire matrix contains all the information between the angular correlation properties of scattering potentials of between particles within one type and across different types we may call it pair-potential matrix (PPM) \cite{DDY}, which has never been noticed before.

The second matrix is defined as
\begin{align}\label{matrix2}
    \mathcal{G}(\mathbf{K_{1}},\mathbf{K_{2}},\omega)&=\Bigl[\widetilde{C}_{g_{pq}}(-\mathbf{K_{1}},\mathbf{K_{2}},\omega)\Bigr]_{\mathcal{L}\times\mathcal{L}}.
\end{align}
From Eqs. \eqref{densitycorrelation} and \eqref{3}, it also readily follows that the diagonal elements of this matrix stand for angular self-correlations of the density distributions of particles of same type and the off-diagonal elements stand for angular cross-correlations of the density distributions of each pair of particle types. The entire matrix contains all the information between the angular correlation properties of density distributions between particles within one type and across different types. Notice that $\mathcal{G}(\mathbf{K_{1}},\mathbf{K_{2}},\omega)$ is essentially same as the pair-structure matrix introduced by Tong $\textit{et al}$ before \cite{Tong}, and one can see from here that the elements of the PSM, i.e., the pair-structure factors and joint pair-structure factors, are in fact closely related to the angular correlations of density distributions of between particles within one type and across different types in the collection.

One may also need to notice that $\mathcal{F}(\mathbf{K_{1}},\mathbf{K_{2}},\omega)$ and $\mathcal{G}(\mathbf{K_{1}},\mathbf{K_{2}},\omega)$ are, in general, not Hermitian matrices since $\widetilde{C}_{{f}_{qp}}(-\mathbf{K_{2}},\mathbf{K_{1}},\omega)\neq\widetilde{C}_{f_{pq}}^{*}(-\mathbf{K_{1}},\mathbf{K_{2}},\omega)$ and $\widetilde{C}_{g_{qp}}(-\mathbf{K_{2}},\mathbf{K_{1}},\omega)\neq\widetilde{C}_{g_{pq}}^{*}(-\mathbf{K_{1}},\mathbf{K_{2}},\omega)$. However, in many situations of practical interest, i.e., their elements have common Gaussian Schell-model distributions \cite{Dingz} and multi-Gaussian Schell-model distributions \cite{ZJZ} as well as quasi-homogeneous distributions \cite{CW}, $\mathcal{F}(\mathbf{K_{1}},\mathbf{K_{2}},\omega)$ and $\mathcal{G}(\mathbf{K_{1}},\mathbf{K_{2}},\omega)$ are symmetric with respect to $\mathbf{K}_{1}$ and $\mathbf{K}_{2}$, and thus they can be Hermitian.

With these two matrices in hands and making use of the well-known trace operation of matrix together with Eq. \eqref{crossspectraldensity1}, the cross-spectral density of the scattered field can be reformulated as
\begin{align}\label{crossspectraldensity2}
    W^{(\text{s})}(r\mathbf{s}_{1},r\mathbf{s}_{2},\omega)&\propto\text{Tr}\bigl[\mathcal{F^{^\top}}({{\mathbf{K}}}_{1}{,{\mathbf{K}_{2}},\omega})\cdot \mathcal{G}({{\mathbf{K}}}_{1}{,{\mathbf{K}_{2}},\omega})\bigr],
\end{align}
where $\top$ and $\cdot$ stand for transpose operation and the ordinary multiplication operation, respectively. 

Eq. \eqref{crossspectraldensity2} is one of the main result in this work, which builds a closed-form relation that associates the cross-spectral density function of the scattered field with the PPM and the PSM. It shows that in addition to a trivial factor $S^{\text{(in)}}(\omega)/r^{2}$, the cross-spectral density of the scattered field exactly equals the trace of the product of the PSM and transpose of the PPM, i.e., $\mathcal{F}^{^\top}(\mathbf{K_{1}},\mathbf{K_{2}},\omega)\cdot\mathcal{G}(\mathbf{K_{1}},\mathbf{K_{2}},\omega)$, and thus all the second-order statistical properties of the scattered field can be completely determined from the PPM and the PSM. Clearly, both the PPM and the PSM play an indispensable role in characterizing the weak scattering process of light waves from a hybrid particulate system and neither of them is better than the other. Moreover, Eq. \eqref{crossspectraldensity2} is not limited to whether the the randomness of the scattering potentials of particles or their density distributions in the collection is invoked or not. This is because the PPM and the PSM themselves are independent of such a randomness, unlike the approach in \cite{Tong}, where the validity of the column vector formed by the scattering potentials of all particles is up to this randomness.

We now consider the spectral degree of coherence (SDOC) of the scattered field generated by a collection of particles with $\mathcal{L}$ types, in terms of Eq. \eqref{crossspectraldensity2}. The SDOC can be readily computed from its definition (\cite{Wolf2}, Sec. 4.2) as
\begin{widetext}
\begin{align}\label{SDOC}
 {\mu}^{\text{(s)}}\left( {r{\mathbf{s}}}_{1}{,r{\mathbf{s}_{2}},\omega}%
\right)= \frac{\text{Tr}\bigl[\mathcal{F^{^\top}}({{\mathbf{K}}}_{1}{,{\mathbf{K}_{2}},\omega})\cdot \mathcal{G}({{\mathbf{K}}}_{1}{,{\mathbf{K}_{2}},\omega})\bigr]}{\sqrt{\text{Tr}\bigl[\mathcal{F^{^\top}}({{\mathbf{K}}}_{1}{,{\mathbf{K}_{1}},\omega})\cdot \mathcal{G}({{\mathbf{K}}}_{1}{,{\mathbf{K}_{1}},\omega})\bigr]}\sqrt{\text{Tr}\bigl[\mathcal{F^{^\top}}({{\mathbf{K}}}_{2}{,{\mathbf{K}_{2}},\omega})\cdot \mathcal{G}({{\mathbf{K}}}_{2}{,{\mathbf{K}_{2}},\omega})\bigr]}}.   
\end{align}
\end{widetext}

Eq. \eqref{SDOC} is the final expression to show how the SDOC of the scattered field depends on the angular correlation properties of the whole collection, which include two parts: one has to do with the angular correlations of the scattering potentials of particles within one type and cross different types, and the other with angular correlations of the density distributions of particles within one type and cross different types. We now show that Eq. \eqref{SDOC} has a pretty interesting result in a special situation, i.e., for a situation where the spatial distributions of scattering potentials of different types of particles are similar and the same is true of their density distributions, i.e., $\widetilde{C}_{f_{pq}}(-\mathbf{K},\mathbf{K},\omega)\approx\widetilde{C}_{f}(-\mathbf{K},\mathbf{K},\omega)$ and $\widetilde{C}_{g_{pq}}(-\mathbf{K},\mathbf{K},\omega)\approx\widetilde{C}_{g}(-\mathbf{K},\mathbf{K},\omega)$, Eq. \eqref{SDOC} can be simplified as

\begin{equation}\label{SDOC1}
 {\mu}^{\text{(s)}}\left( {r{\mathbf{s}}}_{1}{,r{\mathbf{s}_{2}},\omega}%
\right)=\frac{1}{L^2}\text{Tr}\bigl[\mathcal{U^{^\top}}({{\mathbf{K}}}_{1}{,{\mathbf{K}_{2}},\omega})\cdot \mathcal{G}({{\mathbf{K}}}_{1}{,{\mathbf{K}_{2}},\omega})\bigr],   
\end{equation}
where
\begin{align}\label{matrix22}
    \mathcal{U}(\mathbf{K_{1}},\mathbf{K_{2}},\omega)&=\Bigl[\mathcal{U}_{pq}(\mathbf{K_{1}},\mathbf{K_{2}},\omega)\Bigr]_{\mathcal{L}\times\mathcal{L}}
\end{align}
with
\begin{align}
    \mathcal{U}_{pq}(\mathbf{K_{1}},\mathbf{K_{2}},\omega)=\frac{\widetilde{C}_{f_{pq}}(-\mathbf{K}_{1},\mathbf{K}_{2},\omega)}{\sqrt{\widetilde{C}_{f}(-\mathbf{K}_{1},\mathbf{K}_{1},\omega)}\sqrt{\widetilde{C}_{f}(-\mathbf{K}_{2},\mathbf{K}_{2},\omega)}},
\end{align}
and
\begin{align}\label{matrix4}
    \mathcal{K}(\mathbf{K_{1}},\mathbf{K_{2}},\omega)&=\Bigl[\mathcal{K}_{pq}(\mathbf{K_{1}},\mathbf{K_{2}},\omega)\Bigr]_{\mathcal{L}\times\mathcal{L}}
\end{align}
with
\begin{align}
    \mathcal{K}_{pq}(\mathbf{K_{1}},\mathbf{K_{2}},\omega)=\frac{\widetilde{C}_{g_{pq}}(-\mathbf{K}_{1},\mathbf{K}_{2},\omega)}{\sqrt{\widetilde{C}_{g}(-\mathbf{K}_{1},\mathbf{K}_{1},\omega)}\sqrt{\widetilde{C}_{g}(-\mathbf{K}_{2},\mathbf{K}_{2},\omega)}}.
\end{align}

In comparison with Eq. (\ref{SDOC}), Eq. \eqref{SDOC1} shows that the number of species of particles in this special case appears as a scaled factor to ensure the normalization of the SDOC, and two new matrices appear, i.e., $\mathcal{U}(\mathbf{K_{1}},\mathbf{K_{2}},\omega)$ and $\mathcal{K}(\mathbf{K_{1}},\mathbf{K_{2}},\omega)$. The elements $\mathcal{U}_{pq}(\mathbf{K_{1}},\mathbf{K_{2}},\omega)$ of the former quantify degree of angular cross-correlation of the scattering potentials of particles of different types (if $p\neq q$) or degree of angular self-correlation of the scattering potentials of particles of same type (if $p=q$), and the elements $\mathcal{K}_{pq}(\mathbf{K_{1}},\mathbf{K_{2}},\omega)$ of the latter quantify degree of angular cross-correlation of the density distributions of particles of different types (if $p\neq q$) or degree of angular self-correlation of the density distributions of particles of same type (if $p=q$). The SDOC in this special case is related to the trace of $\mathcal{U}^{^\top}(\mathbf{K_{1}},\mathbf{K_{2}},\omega)\cdot\mathcal{Q}(\mathbf{K_{1}},\mathbf{K_{2}},\omega)$.

\section{Numerical examples}
In the following, we will take two special hybrid particulate systems as examples to illustrate the importance of our new approach.

(i) For the first model we consider a collection of random particles with determinate density distributions. A representative example may be a collection of particles suspended in a atmospheric mass, where irregular fluctuations in temperature and pressure of atmosphere turbulence usually lead the refractive indices of particles in different locations to be different and to be random functions of position space. If these particles move slowly, at least there will be no appreciable changes in their locations during the whole scattering process, and thus their density distributions may be determinate in space (\cite{Wolf2}, Sec. 6.3.1). For simplicity, we consider the situation where only two types of particles are contained in the collection, and assume that both the self-correlation functions of the scattering potentials of particles of same type and the cross-correlation functions of different types have Guassian forms, i.e., 
\begin{small}
\begin{align}\label{distribution}
{C_{f_{pq}}}\left({{{\mathbf{{r_{1}^{\prime}}}}},{{\mathbf{{r_{2}^{\prime}}}}}%
,\omega}\right)&=A_{0}\exp{\Bigl[-\frac{\mathbf{r}_{1}^{\prime^{2}}+\mathbf{r}_{2}^{\prime^{2}}}{4\sigma_{pq}^{2}}\Bigr]}\exp{\Bigl[-\frac{(\mathbf{r}_{1}^{\prime}-\mathbf{r}_{2}^{\prime})^2}{2\eta_{pq}^{2}}\Bigr]}, \notag \\ &\qquad (p,q=1,2)
\end{align}
\end{small}
where $A_{0}$ is a positive real constant, and $\sigma_{pq}$ stands for the effective width of the distribution function of particles of same type (if $p=q$) or of different types (if $p\neq q$), and $\eta_{pq}$ stands for the effective correlation width of the distribution function of particles of same type (if $p=q$) or of different types (if $p\neq q$). 

The self-correlation functions of density distributions of particles of same type and the cross-correlation functions of density distributions of particles of different types have the following forms
\begin{equation}\label{densitydistribution}
{C_{g_{pq}}}\left({{{\mathbf{{r_{1}^{\prime}}}}},{{\mathbf{{r_{2}^{\prime}}}}}%
}\right)=\sum_{m(p)}\delta^{*}(\mathbf{r}_{1}^{\prime}-\mathbf{r}_{pm})\sum_{m(q)}\delta(\mathbf{r}_{q}^{\prime}-\mathbf{r}_{qm}).
\end{equation}

In this case, from Eqs. \eqref{2} and \eqref{3}, the elements of the matrices $\mathcal{F}(\mathbf{K_{1}},\mathbf{K_{2}},\omega)$ and $\mathcal{G}(\mathbf{K_{1}},\mathbf{K_{2}},\omega)$ can be readily calculated as
\begin{align}\label{elements}
    \widetilde{C}_{f_{pq}}(-\mathbf{K}_{1},\mathbf{K}_{2},\omega)&=A_{0}\frac{2^{6}\pi^{3}\sigma_{pq}^{6}\eta_{pq}^{3}}{(4\sigma_{pq}^{2}+\eta_{pq}^{2})^{3/2}}\notag \\ & \times
    \exp{\Bigl[-\frac{\sigma_{pq}^{2}}{2}\bigl(\mathbf{K}_{1}-\mathbf{K}_{2}\bigr)^2}\Bigr]\notag \\ &\times\exp{\Bigl[-\frac{\sigma_{pq}^{2}\eta_{pq}^{2}}{2(4\sigma_{pq}^{2}+\eta_{pq}^{2})}\bigl(\mathbf{K}_{1}+\mathbf{K}_{2}\bigr)^2}\Bigr]
\end{align}
and
\begin{align}\label{elements1}
    \widetilde{C}_{g_{pq}}(-\mathbf{K}_{1},\mathbf{K}_{2},\omega)&=\sum_{m(p)}\exp\Bigl[{i\mathbf{K}_{1}\cdot \mathbf{r}_{pm}}\Bigr]\notag \\ &\times\sum_{m(q)}\exp\Bigl[{-i\mathbf{K}_{2}\cdot \mathbf{r}_{qm}}\Bigr].
\end{align}
Once these matrix elements are known, the SDOC of the scattered field is straightforward from Eq. (\ref{SDOC}).

\begin{figure}[bthp]
\centering
\includegraphics[width=8cm]{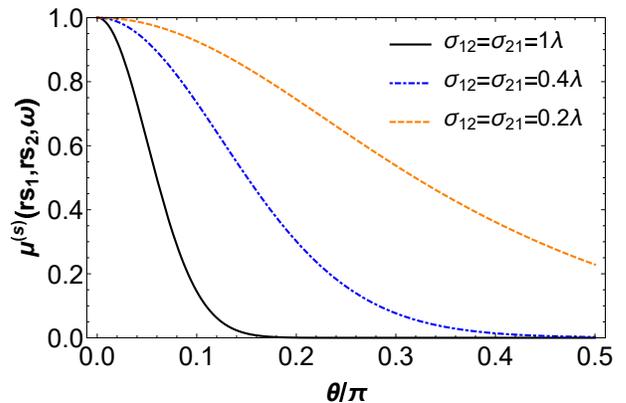}
\caption{Plots of effects of the effective width $\sigma_{12}$ of the correlation function Eq. \eqref{distribution} on the SDOC of light scattering from a collection of random particles with determinate density distributions, as a function of the dimensionless scattering angle $\theta/\pi$. In the current model, we won't pay much attention to the density distributions of particles in the collection, for simplicity, we assume that there are totally four particles in the collection, and each kind has two particles. The coordinates are set to be $(0,0.1\lambda,0)$ and $(0,-0.1\lambda,0)$ for the first kind of particles, and $(0,0.2\lambda,0)$ and $(0,-0.2\lambda,0)$ for the second kind of particles. $\mathbf{s}_{1}=(\sin{\theta_{1}\sin{\phi_{1}}},\sin{\theta_{1}\cos{\phi_{1}}},\cos{\theta_{1}})$,  $\mathbf{s}_{2}=(\sin{\theta\sin{\phi}},\sin{\theta\cos{\phi}},\cos{\theta})$. The parameters for calculations are $\phi=\pi/2$, $\theta_{1}=0$, $\phi_{1}=\pi/2$, $\protect\sigma _{11}=\protect\sigma _{22}=0.1\lambda$, $\protect\eta _{11}=\protect\eta _{22}=0.01\lambda$, $\protect\eta _{12}=\protect\eta _{21}=0.03\lambda$.}
\label{Fig 2}
\end{figure}

\begin{figure}[bthp]
\centering
\includegraphics[width=8cm]{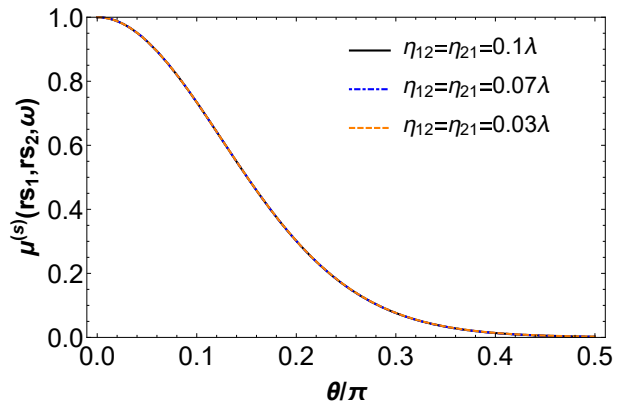}
\caption{Plots of effects of effective correlation width $\eta_{12}$ of correlation function Eq. \eqref{distribution} on the SDOC of light scattering from a collection of random particles with determinate density distributions, as a function of the dimensionless scattering polar angle $\theta/\pi$. $\protect\sigma _{12}=\protect\sigma _{21}=0.4\lambda$, and the other parameters for calculations are the same as Fig. \ref{Fig 2}.}
\label{Fig 3}
\end{figure}

We are mainly interested in how the cross-correlations between the scattering potentials of different types of particles effect the coherence of the scattered field, which has always been neglected before. Fig. \ref{Fig 2} depicts the behaviors of the normalized SDOC of the scattered field for different effective widths $\sigma_{12}$. It is found that when the effective correlation width $\sigma_{12}$ decreases the SDOC of the scattered fields can be enhanced greatly, which means that even if the cross-correlations between the scattering potentials of different types of particles are weak, it can still effect the SDOC strongly. In comparison to the effective correlation width $\eta_{12}$, the effective correlation width $\eta_{12}$ can have a negligible influence on the SDOC of the scattered field, as can be seen from Fig. \ref{Fig 3}. This is because the effective width $\sigma_{12}$ and the effective correlation width $\eta_{12}$ meet the relation $\sigma_{12}/\eta_{12}\gg 1$ in our numerical calculations. In this case, the Gaussian Schell-model distribution in Eq. \eqref{distribution} can reduce to a quasihomogeneous distribution, in which the well-known reciprocity relation holds \cite{TD}, leading the effective correlation width to have no consequence on the SDOC of the scattered field.


(ii) We now restrict our attention to the second model, i.e., a collection of determinate particles with random density distributions. A representative example is a tenuous collection of cells suspended in a solution, where the refractive indices of different cells are well-defined functions in space, but their density distributions in the solution may be random (\cite{Wolf2}, Sec. 6.3.1). The current model has been taken into account preliminarily \cite{Tong}, where use has been made of the assumption that the density distributions of different types of particles are similarly distributed in space. Here we will relax this constraint, and 
concern ourselves with a more general case, viz.,
\begin{align}\label{distribution2}
{C_{g_{pq}}}\left({{{\mathbf{{r_{1}^{\prime}}}}},{{\mathbf{{r_{2}^{\prime}}}}}}\right)&=C_{0}\exp{\Bigl[-\frac{\mathbf{r}_{1}^{\prime^{2}}+\mathbf{r}_{2}^{\prime^{2}}}{4\gamma_{pq}^{2}}\Bigr]}\exp{\Bigl[-\frac{(\mathbf{r}_{1}^{\prime}-\mathbf{r}_{2}^{\prime})^2}{2\delta_{pq}^{2}}\Bigr]}, \notag \\ &\qquad (p,q=1,2)
\end{align}
where $C_{0}$ is also a positive real constant, and $\gamma_{pq}$ and $\delta_{pq}$ have the same meaning as $\sigma_{pq}$ and $\eta_{pq}$, respectively. Similarly, we still consider the situation where only two types of particles are contained in the collection and both the self-correlation functions of the density distributions of particles of same type and the cross-correlation functions of different types also obey Gaussian distributions. 

The self-correlation functions of the scattering potentials of particles of same type and the cross-correlation functions of different types now have forms
\begin{equation}\label{densitydistribution1}
{C_{f_{pq}}}\left({{{\mathbf{{r_{1}^{\prime}}}}},{{\mathbf{{r_{2}^{\prime}}}}}%
,\omega}\right)= f_{p}^{*}(\mathbf{r_{1}}^{\prime},\omega)f_{q}(\mathbf{r_{2}}^{\prime},\omega).
\end{equation}
where
\begin{equation}\label{scatteringpotential2}
f_{p}(\mathbf{r}^{\prime},\omega)=B_{0}\exp{\Bigl(-\frac{\mathbf{r}^{\prime 2}}{2\zeta_{p}^{2}}\Bigr)}
\end{equation}
is the scattering potential of the $p$th-type particle, with $B_{0}$ being a positive real constant. 
From Eqs. \eqref{2} and \eqref{3}, it is now seen that the elements of the matrices $\mathcal{F}(\mathbf{K_{1}},\mathbf{K_{2}},\omega)$ and $\mathcal{G}(\mathbf{K_{1}},\mathbf{K_{2}},\omega)$ have become
\begin{align}\label{elements11}
    \widetilde{C}_{g_{pq}}(-\mathbf{K}_{1},\mathbf{K}_{2},\omega)&=C_{0}\frac{2^{6}\pi^{3}\gamma_{pq}^{6}\delta_{pq}^{3}}{(4\gamma_{pq}^{2}+\delta_{pq}^{2})^{3/2}}\notag \\ & \times\exp{\Bigl[-\frac{\gamma_{pq}^{2}}{2}\bigl(\mathbf{K}_{1}-\mathbf{K}_{2}\bigr)^2}\Bigr]\notag \\ &\times\exp{\Bigl[-\frac{\gamma_{pq}^{2}\delta_{pq}^{2}}{2(4\gamma_{pq}^{2}+\delta_{pq}^{2})}\bigl(\mathbf{K}_{1}+\mathbf{K}_{2}\bigr)^2}\Bigr]
\end{align}
and
\begin{align}\label{elements3}
    \widetilde{C}_{f_{pq}}(-\mathbf{K}_{1},\mathbf{K}_{2},\omega)&=B_{0}^{2}(2\pi\zeta_{p})^6\exp{\Bigl[-\frac{1}{2}\zeta_{p}^{2}\bigl(\mathbf{K}_{1}^{2}+\mathbf{K}_{2}^{2}\bigr)}\Bigr].
\end{align}
With these matrix elements in hands, the SDOC of the scattered field is straightforward from Eq. (\ref{SDOC}) once again.

\begin{figure}[bthp]
\centering
\includegraphics[width=8cm]{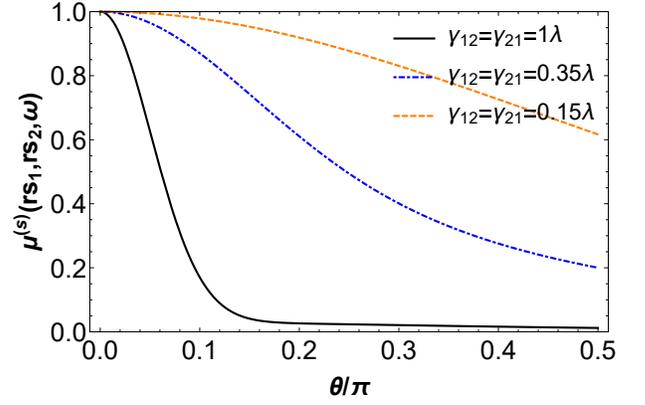}
\caption{Plots of effects of the effective width $\gamma_{12}$ of the correlation function Eq. \eqref{distribution2} on the SDOC of light scattering from a collection of determinate particles with random density distributions, as a function of the dimensionless scattering polar angle $\theta/\pi$. The parameters for
calculations are $\phi=\pi/2$, $\theta_{1}=0$, $\phi_{1}=\pi/2$, $\zeta_{1}=0.2\lambda$, $\zeta_{2}=0.1\lambda$, $\protect\gamma _{11}=\protect\gamma _{22}=0.1\lambda$, $\protect\delta _{11}=\protect\delta _{22}=0.02\lambda$, $\protect\delta _{12}=\protect\delta _{21}=0.01\lambda$.}
\label{Fig 4}
\end{figure}
\begin{figure}[bthp]
\centering
\includegraphics[width=8cm]{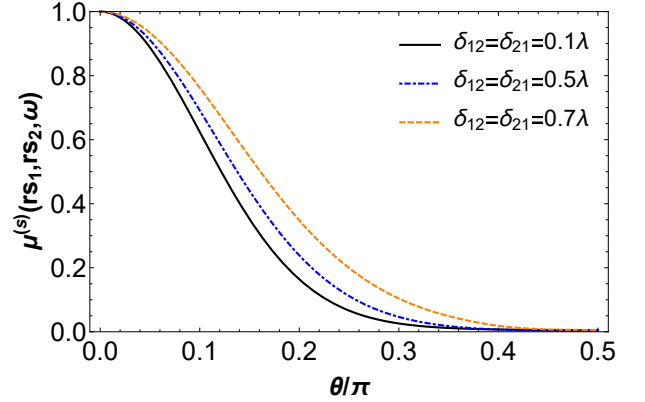}
\caption{Plots of effects of the effective correlation width $\gamma_{12}$ of the correlation function Eq. \eqref{distribution2} on the SDOC of light scattering from a collection of determinate particles with random density distributions, as a function of the dimensionless scattering angle $\theta/\pi$. $\protect\gamma _{11}=\protect\gamma _{22}=0.1\lambda$, $\protect\gamma _{12}=\protect\gamma _{21}=1\lambda$, $\protect\delta _{11}=\protect\delta _{22}=0.2\lambda$. The other parameters for calculations are the same as Fig. \ref{Fig 4}.}
\label{Fig 5}
\end{figure}

Fig. \ref{Fig 4} presents the behaviors of the SDOC of the scattered field for different effective widths $\gamma_{12}$. It is found that when the effective width $\gamma_{12}$ increases the SDOC of the scattered field decreases significantly, and the SDOC of the scattered field in the current model can still has an appreciable value even if the scattered field is observed at large scattering polar angles, for the same magnitude of the effective width as the first model. Fig. \ref{Fig 5} shows the distributions of the SDOC for three different effective correlation widths $\delta_{12}$ of the correlation function, where $\delta_{12}$ is now comparable to $\gamma_{12}$. From Fig. \ref{Fig 5} it follows that the effective angular width of the SDOC increases with the increase of the cross-correlation width of correlation function, which demonstrates intuitively that the influence of the effective correlation width on the SDOC of the scattered field can be observed provided that the constraint $\gamma_{12}/\delta_{12}\gg 1$ is relieved. 

\section{Summary and Discussion}

In summary, we have developed a new approach to the scattering of light waves from a collection of particles with $\mathcal{L}$ types, based on the PPM and the PSM. We derived a closed-form relation that
associates the cross-spectral density function of the scattered field with the PPM and the PSM,  showing that the the cross-spectral density function equals the trace of the product of the PSM and the transpose of the PPM. This means that these two matrices are not only sufficient to determine all the second-order statistical properties of the scattered field, but also can largely simplify the theoretical procedures of the scattering of light from complex collection of scatters. Based on this, the spectral degree of coherence (SDOC) of the scattered field was further analysed, and we showed that for a special case where the spatial distributions of scattering potentials of different types of particles are similar and the same is true of their density distributions, the PPM and the PSM will reduce to two new matrices whose elements separately quantify degree of angular correlation of the scattering potentials of particles and their density distributions, and the number of species of particles as a scaled factor ensures the normalization of the SDOC. Two hybrid particulate systems as examples illustrated the influence of the off-diagonal elements of the PPM and the PSM on the SDOC of the scattered field, respectively. Our new approach is compared with the existing solution to waves scattering from a collection of particles with $\mathcal{L}$ types and is found superior to the previous attempt, and it is therefore expected to have applications in such scientific and engineering communities as atmosphere optics and biomedical diagnostics where collections of particles of different types are often encountered.

\section{Acknowledgement}

The author acknowledges Zhenfei Jiang at Texas A $\&$ M University for her passionate assistance during the author visit to University of Rochester. Financial support was provided by Fundamental Research Funds for the Central Universities No. 2682022CX040.


\begin{thebibliography}{99}
\bibitem{DDY} The PPM here is different from the pair-scattering matrix introduced in the context of $S$-matrix theory \cite{Skoro}. The latter is a generalized matrix which transforms the angular correlation function of the incident field into the angular correlation function of the scattered field, and it is not difficult to imagine from our current method that the pair-scattering matrix can also be related to the trace of the product of a $\mathcal{L}\times\mathcal{L}$ PPM and a $\mathcal{L}\times\mathcal{L}$ PSM, which will be discussed in detail elsewhere.
\end{thebibliography}
\end{document}